\documentclass[aps,preprint,onecolumn,citeautoscript,draft,superscriptaddress]{revtex4-1} 
\usepackage{float}
\usepackage{amsmath,amssymb,mathrsfs,bm,feynmf,setspace}
\usepackage{graphicx}
\usepackage[tight]{subfigure}  
\usepackage{color}
\usepackage{verbatim} 
\usepackage[papersize={8.5in,11in}]{geometry}
\usepackage[colorlinks=true]{hyperref} 
\hypersetup{
    bookmarks=true,         
    unicode=false,          
    pdftoolbar=true,        
    pdfmenubar=true,        
    pdffitwindow=false,     
    pdfstartview={FitH},    
    pdftitle={My title},    
    pdfauthor={Author},     
    pdfsubject={Subject},   
    pdfcreator={Creator},   
    pdfproducer={Producer}, 
    pdfkeywords={keyword1} {key2} {key3}, 
    pdfnewwindow=true,      
    colorlinks=true,       
    linkcolor=red,          
    citecolor=blue,        
    filecolor=magenta,      
    urlcolor=cyan           
} 
\usepackage{bm}
\newcommand{\be}{\begin{equation}}
\newcommand{\ee}{\end{equation}}
\newcommand{\bea}{\begin{eqnarray}}
\newcommand{\eea}{\end{eqnarray}}

\def\bse{\begin{subequations}}
\def\ese{\end{subequations}}

\newcommand{\el}[1]{\label{#1}}
\newcommand{\er}[1]{\eqref{#1}}

\newcommand{\ci}[1]{}

\newcommand{\ke}{\rangle}
\newcommand{\br}{\langle}
\newcommand{\lb}{\left(}
\newcommand{\rb}{\right)}
\newcommand{\lc}{\left.}
\newcommand{\rc}{\right.}
\newcommand{\lsb}{\left[}
\newcommand{\rsb}{\right]}

\newcommand{\nn}{\nonumber \\}
\newcommand{\p}{\partial}

\newcommand{\eq}[1]{(\ref{#1})}

\newcommand{\ba}{\begin{eqnarray}}
\newcommand{\ea}{\end{eqnarray}}
\newcommand{\bal}{\begin{align}}
\newcommand{\eal}{\end{align}}
\newcommand{\bay}[1]{\left(\begin{array}{#1}}
\newcommand{\eay}{\end{array}\right)}
\newcommand{\eg}{\textit{e.g.} }
\newcommand{\ie}{\textrm{i.e.}, }

\newcommand{\at}[1]{{\Big|}_{#1}}
\newcommand{\zt}[1]{\textrm{#1}}

\def\rmd{{\rm d}}
\def\xa{{\alpha}}

\def\xb{{\beta}}

\def\xd{{\delta}}

\def\xe{{\epsilon}}

\def\xs{{\sigma}}
\def\xS{{\Sigma}}

\def\CO{{\cal O}}

\begin{document}

\title{OPE of the stress tensors and surface operators}
\author{Xing Huang} \email{xingavatar@gmail.com}
\affiliation{Department of Physics, National Taiwan Normal University, Taipei, 116, Taiwan}
 \author{ Ling-Yan Hung} \email{lyhung@fudan.edu.cn}
\affiliation{Department of Physics and Center for Field Theory and Particle Physics, Fudan University, Shanghai 200433, China}
\affiliation{Collaborative Innovation Center of Advanced Microstructures, Fudan University, Shanghai, 200433, China.}
 \author{Feng-Li Lin} \email{linfengli@phy.ntnu.edu.tw} 
 \affiliation{Department of Physics, National Taiwan Normal University, Taipei, 116, Taiwan}

 \date{\today\\
 \vspace{1.6in}}
\begin{abstract}
We demonstrate that the divergent terms in the OPE of a stress tensor and a surface operator of general shape cannot be constructed only from local geometric data depending only on the shape of the surface. We verify this holographically at $d=3$ for Wilson line operators or equivalently the twist operator corresponding to computing the entanglement entropy using the Ryu-Takayanagi formula. We discuss possible implications of this result. 
\end{abstract}
\maketitle

\section{Introduction }
Global symmetries and the Ward identities that follow are extremely important in field theories. It gives stringent constraints to 
the allowed structure of correlation functions. In turn, given knowledge of how fields should transform under a symmetry transformation, this information has to be stored in the OPE coefficients between the global symmetry charge operators and the field operator under consideration. These OPE coefficients  between charge operators and local operators have been studied extensively. To take the simplest example, in a 1+1 dimensional CFT, the OPE between the stress tensor and a primary scalar operator takes the following form:

\be
T_{zz}(z) \mathcal{O}_{\Delta}(0) \sim \frac{\Delta\, \mathcal{O}_\Delta}{z^2} +  \frac{\partial\mathcal{O}_\Delta }{z} + \cdots
\ee
where $\Delta$ is the conformal dimension of the operator $\mathcal{O}_\Delta$.  The first term stores data of how the scalar operator transforms under scaling transformation, while the second term  describes how  the scalar operator transforms under translation.
One naturally wonders how the story generalizes to OPE's of charges with surface operators.  For starter, it is certainly the case that Ward identities still apply whenever there is a global symmetry. If a global symmetry acts locally, it is natural to expect that  a local charge density operator can be defined, and that the OPE coefficients -- in this case coefficients of terms divergent in the inverse of the perpendicular distance of the charge density operator from the surface operator -- should again carry information about how the surface operator transforms under the corresponding symmetry. These considerations should also have generalizations in studying higher symmetries \cite{Gaiotto:2014kfa} which necessarily involve symmetry transformation of higher dimensional objects. OPE coefficients between the stress tensor and a surface operator is a very good starting point to study these questions, particularly so in a CFT theory. The stress tensor exists in any CFT and their expectation values are very strongly constrained by conformal symmetry, which allows us to make guesses of the form of these coefficients. Moreover, for co-dimension two operators wrapping spherical regions for example, the form of the expectation values of the stress tensor is constrained down to only a single  coefficient, thus allowing us to make quantitative checks of our guesses.   

Co-dimension two operators are also very interesting for their connections to our understanding of entanglement entropy. The entanglement entropy and its cousins --the R\'enyi entropy-- hold important information about the underlying field theory. It is known that the  value of the entanglement entropy can be understood as the expectation value of a twist operator that introduces branch cuts over the entangling surface surrounding the region under consideration \cite{cardy0, Casini:2010nn, Swingle:2010jz}.  An understanding of these OPE coefficients of the stress tensor with the twist operators would allow us to compute at ease the entanglement entropy with entangling surfaces deformed arbitrarily from better understood ones, such as planar and spherical entangling surfaces. Ward identities of surface operators, particularly those involving entanglement entropy, have been considered before\cite{Banerjee:2011mg,Billo:2013jda}. More recently there is a flurry of work towards understanding the entanglement entropy of more general shapes at least perturbatively from spherical/planar ones \cite{Rosenhaus:2014woa,Rosenhaus:2014nha, Rosenhaus:2014ula,Rosenhaus:2014zza,Allais:2014ata,Mezei:2014zla}. Our work is the first attempt towards understanding these perturbative changes via OPE coefficients.   
      
Our strategy can be summarized as follows. We insert the stress tensor at a perpendicular distance $\epsilon$ from the surface operator. As in \cite{Berenstein:1998ij,Gomis:2009xg} we would like to focus on the ratio of the expectation value of the stress tensor   to that of the surface operator alone. Simply from dimensional analysis, it is expected that the leading term of the OPE contains a $1/\epsilon^d$ divergence as $\epsilon \to 0$, and that the form of the leading term is controlled by a single coefficient for a conformal field theory\cite{Kapustin:2005py,Hung:2011nu,Hung:2014npa}. It is known that the leading term does not actually take part in controlling the transformation of the surface operator under scaling or translation\cite{Kapustin:2005py}. This follows from Gauss law when we consider integrating the stress tensor over appropriate Gauss surface wrapping the surface operator. We are therefore most interested in the sub-leading terms in $\epsilon$. We make the working assumption that the the diverging sub-leading terms can in fact be constructed using local geometric data--such as extrinsic curvatures and torsion-- of the surface on which the operator wraps. This approach was taken in \cite{Azuma:2001hc} where the authors there attempted a systematic study of OPE's of Wilson line operators in four dimensions, which is the classic example of extended operators. For simplicity, we focus on $d=3$ theories, so that a co-dimension 2 operator is essentially a line operator, much like the Wilson line operator. We demonstrate however that our working assumption necessarily breaks down-- that it is not possible to satisfy stress energy conservation if we insist upon constructing the sub-leading terms out of local geometric quantities. We check our claims by comparing our results with a holographic calculation of the stress tensor in the presence of a surface operator by inserting probes in the bulk. The holographic calculation produces an expectation of the stress tensor that is consistent with stress energy conservation. We demonstrate however that it is not expressible in terms of local geometric data, giving support to our field theoretic analysis. 

An outline of our discussion is as follows: we begin a review of the Ward identities of extended objects in section \ref{Ward}, and then work out a systematic expansion of the expectation value of the stress tensor in the vicinity of a line operator in 2+1 d as a power series expansion in the infinitesimal perpendicular distance $\epsilon$, assuming that the terms are local geometric covariants. We obtain constraints of the coefficients of these terms up to the first subleading order in $\epsilon$ using tracelessness and stress energy conservation. In section \ref{sec3}, we compare these constraints with the known results of a line operator wrapping a circle, and which we find complete agreement, although we demonstrate why this program would begin to fail in the next sub-leading order in the $\epsilon$ expansion.
In section \ref{holo}, we compute the stress tensor induced by the presence of a line operator slightly deformed by a transverse wave from a straight line holographically, and find evidence that support our claim of a non-local stress tensor expansion. We conclude our paper with a few final comments in section \ref{conclude}. Some details of the calculations are relegated to the appendices.

\section{Ward Identities and OPEs}\label{Ward}
As shown in \cite{Banerjee:2011mg}, under diffeomorphism, the entanglement entropy which is a special instance of a co-dimension 2 operator insertion, changes as follows:
\be
S_{EE}(A,f*g)- S_{EE}(A,g) = \int dx^{n+1} \sqrt{g} \xi_\mu \nabla_{\nu}\left( -\frac{2}{\sqrt{g}} \frac{\delta S_{EE}}{\delta_{g_{\mu\nu}}}\right)
\ee
where $f$ is a time-independent spatial diffeomorphsim continuously connected to the identity, and $A$ denotes the region within the entangling surface on which the twist operator is inserted. 
Since 
\be
S_{EE}(f(A),g) = S_{EE}(A,f*g),
\ee
where $f*g$ denotes the pull back of the metric, one can replace a smooth deformation of the entangling surface by diffeomorphism.

The above is equivalent to
\be
 \nabla_{\nu}\left( -\frac{2}{\sqrt{g}} \frac{\delta S_{EE}}{\delta{g_{\mu\nu}}}\right) = F n^\mu  \delta_{\Sigma},
\label{conserveEq}
\ee
where $\Sigma$ is the entangling surface, and $F$ is some general function which depends on on coordinates tangential to $\Sigma$, and $n^\mu$ the unit normal vector, normal to the entangling surface and in the radial direction of the codimension 2 plane. This is the statement that the twist operator inserted over $\Sigma$ breaks energy-momentum conservation, and momentum is only unconserved in the orthogonal direction.

The l.h.s. is basically just the expectation value of the stress tensor in the presence of the twist operator.
We would like to extract $F$, in which case we would be able to deal with any arbitrary deformation of the entangling surface, given the expectation value of the stress tensor.

\subsection{Constructing expectation value of the stress tensor relating to OPE }
\label{IIA}

From the discussion above, it is clear that at the end of the day, it is the UV divergent terms in the stress tensor expectation value that carries information about the functions $F$. Therefore, to extract this function for more general entangling surfaces, we would like to expand the stress tensor in powers of $|\epsilon|$. The expansion coefficients are then related to OPE coefficients of stress tensor with the surface operator inserted at the entangling surface.

To simplify the discussion, we will focus on $d=3$, so that the entangling surface would be one dimensional. 

The expansion of the stress tensor should take the following form,
\be
T_{\mu\nu}(x) = \frac{1}{|\epsilon|^3} \bigg( A^0_{\mu\nu} + |\epsilon| A^1_{\mu\nu} + \cdots \bigg)\;. 
\ee
We assume the OPE coefficients $A^i_{\mu\nu}$ can be expressed in terms of local geometry data. Thus, for the leading term the most general form should be as follows: 
\be
A^0_{\mu\nu} = a_1 \delta_{\mu\nu} + a_2 t_\mu t_\nu + a_3 \hat{n}_\mu \hat{n}_\nu,
\ee
where $t^\mu(\hat{x})$ is the unit tangent vector on the entangling surface evaluated at the point $\hat{x}$, and $\hat{n}$ is again the unit normal vector defined by 
\be
\hat{n}_\mu =\frac{ (x-\hat{x})_\mu}{\epsilon}.
\ee 
Some useful identities regarding the local geometrical quantities are relegated into the appendix \ref{appendixA}.

Similarly, by using the dimensional analysis and parity symmetry, we can build $A^1_{\mu\nu}$ out of the local geometric data as follows
\be
A^1_{\mu\nu} = K^1 (b_1 \delta_{\mu\nu} + b_2 t_\mu t_\nu + b_3 \hat{n}_\mu\hat{n}_\nu) +b_4 (\hat{n}_\mu n^a_\nu K^a + \hat{n}_\nu n^a_\mu K^a).
\ee
We note that the rule is that $\hat{n}$ and its derivatives thereof are allowed to appear an odd number of times at each order of the expansion, whereas $t^\mu$ and $k$ must appear an even number of times, to preserve the symmetry of the problem.

To continue, we also have
\bea
\label{A2-express}
A^2_{\mu\nu} = && c_0 \partial_\sigma K^1 n^1_{(\mu} t_{\nu)} + c_0' \partial_\sigma K^a n^a_{(\mu} t_{\nu)} 
+ (K^1)^2 \left( c_1 \delta_{\mu\nu} +c_2 t_\mu t_\nu + c_3 n^1_\mu n^1_\nu\right) + \nonumber \\
&& + c_4 K^1 ( n^1_\mu n^a_\nu K^a + n^1_\nu n^a_\mu K^a ) + c_5  n^a_\mu n^b_\nu K^a K^b  \nonumber \\
&& + (K^a)^2\left(c''_1 \delta_{\mu\nu} +c''_2 t_\mu t_\nu + c''_3 \hat{n}_\mu\hat{n}_\nu\right) \,.
\eea

We note that these parameters, $\{a_i,b_i,c_i, \cdots\}$ are not independent, since the stress tensor is conserved anywhere away from the entangling surface, and moreover in a conformal theory it has to be traceless. We can reduce the number of independent coefficients using these constraints, order by order in perturbation theory.

  Plug in the expansion of $T_{\mu\nu}$ one can obtain
\be\label{partialT}
\partial^\mu T_{\mu\nu} = \frac{-3}{|\epsilon|^{4}} n^{1\,\mu} \left( A^0_{\mu\nu} + |\epsilon| A^1_{\mu\nu} + \cdots \right) + \frac{1}{|\epsilon|^3} \left(\partial^\mu A^0_{\mu\nu} + n^{1\, \mu} A^{1}_{\mu\nu} + |\epsilon| \partial^\mu A^1_{\mu\nu} + \cdots \right)\;.
\ee

Furthermore, 
\be
\partial^\mu A^0_{\mu\nu} =  a_2\frac{n^1_\nu K^a}{1-|\epsilon|K^1}+ \frac{a_3 n^1_\nu (2- \frac{1}{(1-|\epsilon| K^1)})}{|\epsilon|},
\ee
and that 
\bea
\label{conservA1}
\partial^\mu A^1_{\mu\nu} =&& (\partial_\sigma K^1 - K^a n^a_\beta \partial_\sigma n^{1\, \beta}) (b_1+b_2) s_\nu +\frac 1 {|\xe|} b_1 (K^an^a_\nu-K^1 n^1_\nu)\nonumber \\
&& + K^1 (\frac{b_3 n^1_\nu}{|\epsilon|} (2 - \frac{1}{1- |\epsilon|K^1}) + b_2 \frac{n^a_\nu K^a}{1-|\epsilon|K^1}) \nonumber \\
&&+ b_4 \left( \frac{n^a_\nu K^a}{|\epsilon|} (2 - \frac{1}{1-|\epsilon|K^1}) + \frac{K^a}{|\epsilon|} (n^a_\nu - \delta^{a1}n^1_\nu) - t_\alpha s^\alpha n^1_\nu (K^a)^2 )\right)
\eea
where $s_{\mu}={t_{\mu} \over 1-|\epsilon| K^1}$ as defined in \eq{dn1}.

Now collecting all the results and require  energy-momentum conservation $\partial^\mu T_{\mu\nu} =0$, at the leading order of $\epsilon$-expansion of \eq{partialT}, i.e., $\mathcal{O}(|\epsilon|^{-4})$, we have
\be\label{cons-1}
3 a_1 + 2 a_3 =0.
\ee

To first sub-leading order, there are two independent terms whose coefficients we set to zero:
\be\label{cons-2}
-n^1_\nu K^1 \left(3 b_1+ b_3+ d b_4 + a_3\right) + n^a_\nu K^a (a_2+ b_1) =0\,,
\ee
giving
\be\label{cons-3}
3 b_1+ b_3+ 3 b_4 + a_3=0 , \qquad a_2+ b_1=0.
\ee

Including constraints from tracelessness, we have in addition
\be
\label{cons-4}
3 a_1 + a_2 + a_3 =0, \qquad 3 b_1+b_2+ b_3 + 2 b_4 =0.
\ee
In the next section we will check these relations by the known result of a spherical twist operator, and then consider the more general surface operators.

\section{Checking the circle in a conformal field theory}\label{sec3}

In a conformal theory, we can determine the stress tensor in the presence of a spherical twist operator inserted along the spherical entangling surface. 
That is simply obtained by conformal transformation from the space $H_{d-1} \times S^1$ via 
\be
\frac{R-\omega}{R+\omega} = \exp(-\sigma), 
\ee
where $\omega = r+ i t$ are the radial and Euclidean time coordinates respectively in the flat frame, and $\sigma = u+ i \tau$ are the hyperbolic radial coordinate in $H_{d-1}$ and Euclidean time coordinate of $S^1$ respectively. 

In the $H_{d-1} \times S^1$ frame, purely from symmetry one can deduce that the stress tensor is diagonal, taking the following form:
\be
T_{\tau\tau} = -(d-1) P, \qquad T_{AB} = P g_{AB},
\ee
where $g_{AB}$ denotes the metric of $H_{d-1}$

Following through the conformal transformation, the stress tensor in the flat frame takes the following form:
\bea
T_{tt} &&= \frac{(2R)^dP}{4|R^2-\omega^2|^d} \left[ 2(2-d) - d (\frac{R^2- \omega^2}{R^2-\bar{\omega}^2} + \frac{R^2-\bar{\omega}^2}{R^2-\omega^2})\right], \\
T_{x_ix_j} &&= \frac{(2R)^d P}{4|R^2-\omega^2|^d} \left[ \left(2(2-d) + d(\frac{R^2- \omega^2}{R^2-\bar{\omega}^2} + \frac{R^2-\bar{\omega}^2}{R^2-\omega^2}) \right) \frac{x_ix_j}{r^2} + 4 (\delta_{ij} - \frac{x_ix_j}{r^2})       \right], \\
T_{tr} &&= i \left\vert\frac{2R}{R^2-\omega^2}\right\vert^d \frac{d P}{4} \left(\frac{R^2-\bar{\omega}^2}{R^2-\omega^2}-\frac{R^2- \omega^2}{R^2-\bar{\omega}^2} \right).
\eea

These expressions can be expanded in $\epsilon$, where we take
\be
\omega = R + (\delta r + i \delta t), \qquad  |\epsilon|^2 = \delta r^2 + \delta t^2,
\ee
from which we can extract the values of $\{a_i,b_i\}$ and so on defined in the previous section
at least for $d=3$. 

A useful expression to make the comparison is that the extrinsic curvatures are given by
\be
K^1 = -\frac{\delta r}{R|\epsilon|}, \qquad  n^a_\mu K^a = -\frac{x_i}{R^2}\bigg\vert_{x^2 = R^2} .
\ee

This allows us to evaluate the expected value of the stress tensor. In particular,
\be
A^1_{tt} = \frac{- \delta r}{R |\epsilon|} (b_1 + b_3 \frac{\delta t^2}{|\epsilon^2|} ), \qquad A^1_{rr} = -\frac{\delta r}{R|\epsilon|} (b_1+ b_3 \frac{\delta r^2}{\epsilon^2} + 2 b_4).
\ee

This can be compared with the corresponding expansion of the stress tensor
\bea
T_{tt}  &&= \frac{P}{|\epsilon|^{d+2}} \left( (\delta t^2 - 2 \delta r^2)  + 3 \delta r \frac{(\delta t^2 +2 \delta r^2)}{2R}\right) + \cdots  \\
T_{rr} &&= \frac{P}{|\epsilon|^{d+2}} \left( (\delta r^2 - 2 \delta t^2)  - 3 \delta r \frac{\delta r^2 }{2R}\right) + \cdots 
\eea

Comparing the two expressions, we get
\bea
\label{coefficientsph}
&& a_1= - 2 P, \qquad a_3 = 3 P = a_2 \\
&&b_1 = -3 P, \qquad b_3 = \frac{3}{2} P  = b_4, \qquad b_2 = \frac{9P}{2}.
\eea
We note that all the constraints \er{cons-1} \er{cons-3} \er{cons-4} following from tracelessness and momentum conservation are satisfied as expected. We note also that these coefficients are over-determined and so provides a non-trivial check of the current construction.

\subsection{Hitting a rock--- Failing Ward identity by local OPE for general entangling surfaces}
Despite that we have some success applying the ansatz to spherical entangling surface, challenge comes immediately when we try to deal with less special surface $\xS$. The stress tensor from the ansatz is not conserved, nor does it provide the correct entanglement entropy perturbatively.
\subsubsection{Evidence 1}

   In \ref{IIA} we consider the energy-momentum conservation up to $\mathcal{O}(|\epsilon|^{-3}$ and it yields the constraints \eq{cons-1}-\eq{cons-3}. Now we will see that the conservation cannot hold at the order $\mathcal{O}(|\epsilon|^{-2})$.  At this order, the terms contributed to $\partial^\mu T_{\mu\nu}$ are from the order $\mathcal{O}(|\epsilon|^0)$ terms of 
\be
\partial^{\mu} A^1_{\mu\nu}-n^{1\mu} A^2_{\mu} + |\epsilon| \partial^{\mu} A^2_{\mu\nu}\;.
\ee

As we can see from \eq{conservA1}, the term at this order from $\partial^\mu A^1_{\mu\nu}$ is ,
\be\label{leading-c}
\partial^\mu A^1_{\mu\nu}|_{\mathcal{O}(|\epsilon|^0)} =(b_1+b_2) s_\nu \partial_{\sigma} K^1  + \mathcal{O}(K^a)^2
\ee
where the first term is proportional to $t_{\nu}$ and is of the first order of $\partial_{\sigma} K^1$, and the other terms involve $(K^a)^2$ but not $\partial_{\sigma} K^1$.   On the other hand, from the defining equation \eq{A2-express} the terms in $n^{1\mu} A^2_{\mu\nu}|_{\mathcal{O}(|\epsilon|^0)}$ involving $\partial_{\sigma} K^1$ are instead in the form of $n^{1\mu} \partial_{\sigma} K^1 $ which cannot compensate the term in \eq{leading-c}.

We note that in principle $|\epsilon| \partial^\mu A^2_{\mu\nu}|_{\mathcal{O}(|\epsilon|^0)}$ can also provide terms proportional to $\partial_{\sigma} K^1 t_\mu$.  However as we can show explicitly that 
\ba
|\epsilon| \p^\mu A^2_{\mu\nu} |_{\mathcal{O}(|\epsilon|^0)} &= &  2 c_1 K^1( K^an^a_\nu-K^1 n^1_\nu) + (K^1)^2( c_3 n^1_\nu  + c_2 n^a_\nu K^a) \nonumber \\
&+&  c_4 K^1 \left[ 2 n^a_\nu K^a - n^1_\nu K^1 - t_\alpha s^\alpha n^1_\nu (K^a)^2\right] + c_4[(K^a)^2 -(K^1)^2]  n^1_\nu \nonumber \\
&+ & 2 c_1'' K^1 (K^an^a_\nu-K^1 n^1_\nu)- 2 c_1''K^1 (K^a n^a_\nu - K^1 n^1_\nu) \nn
& + &  (K^a)^2 ( c_3'' n^1_{\nu}  + c_2''  n^a_{\nu} K^a ) \,, 
\ea
where we have used the identities in appendix \ref{appendixA}.  We see no term in the form of $\partial_{\sigma} K^1  t_{\mu}$, which thus cannot be used to cancel the term in \eq{leading-c}.  Moreover, after detailed check we find that the other terms not involving $t_{\mu}\partial_{\sigma}K^1$ will all be cancelled out exactly at this order.


As a result,  the conservation of stress tensor at this order at least requires
\be b_1 + b_2 = 0\,,\ee
which we know is inconsistent with $b_1, b_2$ obtained from the case of spherical entangling surface \er{coefficientsph}. In other words, the stress tensor that follows from our ansatz is not conserved unless the entangling surface satisfies $\p_\xs K^1 = 0$.

\subsubsection{Evidence 2}
 In \cite{Allais:2014ata}, it was shown that the entanglement entropy due to the deformation of a circular entangling surface (at $d=3$) does not receive linear correction from the deformation. Instead, the quadratic correction is non-zero and takes the form
\be
\Delta S_{EE} \sim \epsilon^2 n(n^2-1) (a_n^2 + b_n^2),
\ee
for the deformation in the form of
\be
\delta r \sim \epsilon( a_n \cos n\theta + b_n \cos n\theta)
\ee
where $(r,\theta)$ is the polar coordinate parametrizing the disk region bounded by the entangling circle. 

Similar contributions can be done for the deformation of a flat entangling surface. As we will show from the holographic calculations in the next section, the result is 
\be
\delta S_{EE} \sim \epsilon^2 k^3 (a_k^2 + b_k^2), 
\ee
for the deformation taking the form  
\be
\delta y = \epsilon(a_k \cos kx+ b_k\sin kx)
\ee
to an entangling surface aligned along the $x$ direction.

We note that if this result is obtainable from the OPE of the stress tensor with the twist line operator, it has to come from $A^2_{\mu\nu}$ as the linear order correction from $A^1_{\mu\nu}$ vanishes. Namely, consider an entangling surface only slightly deformed from a flat entangling surface. The deformation is given precisely as a wave above. In this case, the extrinsic curvature of this surface must take the form
\be
K \sim \epsilon k^2 (a_k \cos kx + b_k \sin kx ).
\ee
Now if we further perturb by another $\delta y$ of exactly the same waveform, then to quadratic order in $\epsilon$, the only term that can contribute must be terms involving $\partial_x K \sim \epsilon k^2 (-a_k \sin kx  + b_k\cos kx )$.

The final form of the surface integral of the stress tensor would have the term 
\be
\int A^2_{\mu\nu} n^\mu \xi^\nu \sim \int dx  \, \partial_x K^1 (\xi \cdot t) + \CO(\xe^3)\,,
\ee
which unfortunately vanishes at the order of $\CO(\epsilon^2)$ because $(\xi \cdot t) \sim \CO(\epsilon^2)$. However, there is no other $\mathcal{O}(\epsilon^2)$ term from $\int A^2_{\mu\nu} n^\mu \xi^\nu$, which is local and dependent only on the geometry of the entangling surface.

This appears to be strong evidence that the OPE coefficients of the stress tensor must contain non-local terms, in addition to the terms we have found, despite the fact that the explicit results of a spherical entangling surface satisfies all the constraints following from our local construction of the divergent terms.

\section{Holographic calculation}\label{holo}
 In this section we will re-consider the issues discussed in \ref{sec3} in the holographic dual. Especially we will derive the holographic stress tensor in the presence of the entangling surface, i.e., the twist line operator at d=3, and find that it is conserved. The result implies that the OPE of the stress tensor with the twist line operator at d=3 cannot be in the local expression of geometric quantities as extrinsic curvature and its derivatives. 

\subsection{Planar entangling surface perturbed}\label{subsec:PESP}

   We first consider the variation of the holographic entanglement entropy due to the deformation of the flat entangling surface at $d=3$. This is an analogy to what has been done in \cite{Allais:2014ata} for the deformation of the  spherical entangling surface. 

    We choose the following metric for the dual $\zt{AdS}_4$.
\be
\rmd s^2 = \frac {L^2} {z^2} (d t^2 + dz^2 + dx^2 + dy^2)
\ee
The entangling surface lies at $y=0$ (also $t=0$) and we consider the following
perturbation
\be\label{y-pert}
\delta y_0 = \xe\, e^{i k x}
\ee
for very small $\epsilon$.

We can take the following ansatz for the corresponding minimal surface
\be \label{y-ansatz}
y(x,z) = y_1(z)  \delta y_0
\ee
by also imposing the following boundary conditions to conform to \eq{y-pert} at $z=0$, i.e.,
\be\label{bcc1}
y_1(0)=1,\qquad y_1(\infty)=0\;.
\ee

The minimal surface is obtained by minimizing the area
\be
\el{areafun}
A = \int \frac{\sqrt{ \left(1+|\partial_x y|^2\right) \left(1+|\partial_z y|^2 \right)-|\partial_x y \partial_z y|^2}}{z^2} d x dz\,,
\ee
where we include the complex conjugate to make the area real. The change of the area of the minimal surface due to the deformation \eq{bcc1} is then given by
\be\label{DeltaA}
\Delta A =\epsilon^2  \int dx dz \;  {(\partial_z y_1)^2 + k^2 y_1^2 \over 2 z^2} 
\ee
from which we get the equation of motion for  $y_1(z)$ 
\begin{align}
z \partial_z^2y_1-2 \partial_z y_1-k^2 z y_1=0\;.
\end{align}

Imposing the boundary condition  \eq{bcc1}, we obtain
the special solution (assuming $k>0$ for simplicity),
\be
\el{minimal1st}
y_1(z) = (k z+1) e^{-k z}\,.
\ee
Substituting \er{minimal1st} into \er{DeltaA}, we get the change of the holographic entanglement entropy
\be
\el{perturbedEEholo}
\Delta S_{EE} \sim \Delta A =  \epsilon^2 \frac{k^2 (k \mu +1) e^{-2k \mu}}{2 \mu} \int  d x = \epsilon^2\lb \frac{k^2}{2 \mu }-\frac{k^3}{2} +\mathcal{O}(\mu)\rb \int  dx\,.
\ee
This result is very similar to (44) in \cite{Allais:2014ata}. Moreover, it is nonzero in contrast to the null result from the OPE of stress tensor with twist operator. 

\subsection{Holographic computation of the stress tensor}

  In the previous consideration we can obtain the change of the holographic entanglement entropy, but did not obtain the corresponding holographic stress tensor in the presence of the twist operator, which is the holographic dual to the Ryu-Takayanagi (RT) action of the minimal surface.  Now we would like to derive such a stress tensor holographically and check its conservation directly.  The way to do it is to introduce a (probing) boundary source metric variation, which then backreacts to the bulk metric. We then evaluate the total on-shell bulk action, including the RT one \cite{Ryu:2006bv} which represents the insertion of the entangling surface on the dual field theory side.  That is, the total bulk action is $S_{total}:=n S_{gr}+(n-1)S_{EE}$ where $S_{gr}$ is the bulk gravity action and $S_{EE}$ is the RT one. Taking the variation of the on-shell action w.r.t. the boundary source metric variation and then setting the source to zero, we will obtain the stress tensor in the presence of the entangling surface inserted as a twist operator, i.e., the OPE of stress tensor with twist operator. 

We will not include the backreaction of the surface operator on the background, presumably because by taking the replica index $n$ to one the dimension of the twist operator approaches zero, or equivalently in the holographic bulk its coupling to the background geometry is suppressed by $n-1$. Since the extremal surface satisfies the equations of motion following from the Ryu-Takayanagi action, to linear order in the metric perturbation, we need only to evaluate the action on-shell neither backreacting the metric perturbation nor the extremal surface.

We choose the Poincare coordinates for the AdS space, and the deformed metric due to the boundary source metric perturbation takes the following form,
\be
ds^2 = \frac{1}{z^2}\lsb (1+h_{zz}) dz^2 + (\eta_{\mu\nu}+ h_{\mu\nu}) dx^\mu dx^\nu +2h_{z\mu} dz dx_{\mu} \rsb
\ee
We are following gauge choice in \cite{Liu:1998bu}, in which 
\ba
h_{ij} & = & c_d \int d^dx' \; {z^{d-2} \over   f^{d}} P_{ijab} \ \xd G_{ab}\,\nn
h_{zi} & = & c_d \frac d {d-1} \int d^dx' \; {z^{d-3} \over   f^{d-1}} B_{iab} \ \xd G_{ab}\,\nn 
h_{zz} & = & -c_d \frac d {d-1} \int d^dx' \; {z^{d-2} \over   f^{d}} C_{ab} \ \xd G_{ab}\,,\ea
where 
\be
f=z^2+\hat{x}_{\mu}\hat{x}^{\mu}\;, \qquad \hat{x}^{\mu}:=x^{\mu}-x'^{\mu}\;,
\ee
and 
\ba
c_d = \frac{\Gamma (d)}{\pi^{\frac d 2} \Gamma \left(\frac{d}{2}\right)},\quad B_{iab} = \frac 1 4 \p_i J_{jk} (\hat x) P_{jkab},\quad C_{ab} = J_{ij}(\hat x)\\
J_{\mu\nu}(\hat x) = \xd_{\mu\nu} - 2 \frac {\hat x_\mu \hat x_\nu}{x^2},\quad P_{ijab} = \frac 1 2 (\xd_{ia}\xd_{jb}+\xd_{ib}\xd_{ja})\,, 
\ea
and $\delta G_{\mu\nu}$ is the (probing) source metric variation at the boundary.
We note that the expectation value of the boundary stress tensor is given by
\be
\langle T^{\mu\nu}\rangle_{\Sigma} = \frac{\xd S_{total}}{ \delta G_{\mu\nu}}\vert_{\delta G_{\mu\nu}\to 0} \simeq  \frac{\xd S_{EE}}{\delta G_{\mu\nu}}\vert_{\delta G_{\mu\nu}\to 0}\;.
\ee
In the above, $\simeq$ means that we take $n\longrightarrow 1$ limit to suppress the backreaction to the bulk metric. 

We will focus on the case in which the entangling surface is almost planar, inserted at $t:=x_1=0$ and $y:=x_d=0$.  In this case the extremal surface is falling straight down, described again by $y=0$ and extended along $z, x_2,\cdots x_{d-1}$.  

  In the following we will consider only $d=3$. We  choose the static gauge such that the induced metric on extremal surface takes the form (up to the terms possibly contributing to $\mathcal{O}(\epsilon)^2$ part of $\Delta S_{EE}$), 
\ba
\el{inducedmetric}
k_{ab}d\sigma^a d\sigma^b &=& \frac{1}{z^2}\Bigg[ \lb 1+ h_{zz} + (\partial_z y)^2 + 2 h_{zy} \partial_z y\rb dz^2 + 2 h_{xz} dx dz \nn
 & & +  \left(1+ h_{xx} + (\partial_x y)^2  + 2 h_{xy} \p_x y  \right)dx^2 \Bigg] + \cdots,
\ea
where we have included derivatives of $y$ w.r.t. the world-volume coordinates for latter convenience. For the strictly planar case, $y=0$ and therefore all terms involving $y$ actually vanishes.

To the lowest order in $\xe$, \ie $\CO(1)$, the stress tensor we get is that for a straight twist operator
\be
\br T_{\mu\nu}  {}^{\CO(1)} \ke = -\frac 2 {8\pi \left(t'^2+y'^2\right)^{3/2}} \lb \delta_{\mu\nu} - \frac 3 2 t_\mu t_\nu - \frac 3 2 \hat{n}_\mu \hat{n}_\nu\rb\,,
\ee
which by itself is conserved and traceless.

The shape dependence of the stress tensor appears at the next leading order. Let us first consider the case in which the only nonvanishing component is $\xd G_{xy}$. The $\CO(\xe)$ order contribution to the (change of the) stress tensor reads ($d=3$, $\hat x = x - x'$)
\ba
\br T_{xy}{}^{\CO(\xe)} \ke & = & \frac {\xd} {\xd G_{xy}} \int dx dz\, \lb \frac{\sqrt{1+h_{zz}+2 i \xe k   [h_{xy}+k z\, (h_{xy}+i h_{zy})]  e^{-k z+i k x} +\mathcal{O}(\xe)^2}}{z^2}-\frac 1 {z^2} \rb\at{\delta G_{xy}\to 0} \nonumber \\
&= &\xe\int dx dz\frac{z e^{i k x-kz} }{\pi ^2 \left({t'}^2+\hat{x}^2+{y'}^2+z^2\right)^5} \left\{
6 \hat x (k z+1)\left({t'}^2+\hat x^2-7 {y'}^2+z^2\right) \rc\nn
& &  +k \left({t'}^2+\hat x^2+{y'}^2+z^2\right) \lsb{t'}^2 (3 k \hat x+2 i k z+2 i)+3 k \hat x^3+2 i \hat x^2 (k z+1)\rc \nn \label{Txy0}
& & \lc\lc +3 k \hat x \left(z^2-{y'}^2\right)+2 i (k z+1) \left({y'}^2+z^2\right)\rsb\right\}\,.
\ea
In the above we have set $y(x,z)$ of the extremal surface into the form of \eq{y-ansatz} and \eq{minimal1st}.  The integration over $x$ in \eq{Txy0} can be done but we only manage to do the $z$ integration after an expansion in momentum $k$. After some tedious calculation, the result takes the following form,
\be
\br T_{xy} {}^{\CO(\xe)} \ke  = \xe e^{i k x'}\lsb \frac{3 i t'^2 k}{8 \pi  \left(t'^2+y'^2\right)^{5/2}}-\frac{i \left(3 t'^2+2 y'^2\right)k^3}{16 \pi  \left(t'^2+y'^2\right)^{3/2}}-\frac{i
k^4}{8 \pi }+\CO(k^5)\rsb\,.
\ee
The series in $k$ above can also be understood as OPE of stress tensor with the twist operator since $\xd \equiv \left(t'^2+y'^2\right)^{1/2}$ is essentially the minimal distance from the point where $T_{\mu\nu}$ locates to the entangling surface ($y=t=0$), i.e., the location of the twist line operator.  

The other components of stress tensor in the presence of the twist operator can be worked out similarly, and the detailed forms are put in the appendix \ref{appendixB}. Based on the detailed form of the stress tensor listed in appendix \ref{appendixB} order by order in $k$-expansion, it is straightforward to show that the stress tensor is traceless and conserved, i.e.,
\be
\br T^{\mu}_{\mu} {}^{\CO(\xe)} \ke=0, \qquad  \partial^{\mu} \br T_{\mu \nu} {}^{\CO(\xe)} \ke=0\;.
\ee
Note that unlike the tracelessness, the conservation of $ \br T_{\mu \nu} {}^{\CO(\xe)} \ke$ is not easy to see without performing the integration along the extremal surface. 

  The particular form of the OPE coefficients seems to rule out the possibility that they be expressed covariantly in terms of local geometric quantities. For example, we can consider the $\CO(k^2)$ term in $\br T_{yy} {}^{\CO(\xe)}\ke$ (see \eq{Tyy0}) and denote the coefficient as $a_{yy}^{(2)}$. Note that the extrinsic curvature $K$ in this case is $K^y = k^2 \xe e^{ik x}$ and none of $n_\mu^{1,2} ,t_\mu$ have any components proportional to $k^2$. In other words, $a_{yy}^{(2)}$ should contain one $K^y$. On the other hand, $a_{yy}^{(2)}$ is also proportional to $y^3$, or rather $(n_y^1)^3$ and does not appear to follow from any 2-tensor built from $n_\mu^{1,2} ,t_\mu$ and $\xd_{\mu\nu}$.

   With $\br T_{\mu\nu}  {}^{\CO(\xe)} \ke$, one can recover the holographic entanglement entropy computed in sec~\ref{subsec:PESP}. More explicitly, we consider perturbation of the entangling surface given by $\xd y = \tilde \xe\, e^{i k x}$ (\ie $\xe \to \xe + \tilde \xe$). With the following coordinate transformation
\be\el{coordtran} y \to y-\tilde \xe e^{i k x}, \qquad z\to z, \qquad x \to x\;, \ee
the entangling surface can be made as a plane ($y = 0$) again. The resulting first order perturbation
to $G_{\mu\nu}$ is
\be \el{metricperturb} \xd G_{xy} = -i\, k \tilde \xe\, e^{i k x}\,, \ee
with all other components of $\delta G_{ij}$ being zero. This perturbation can also be understood as being generated by the vector $\xi$ whose only nonvanishing component is $\xi_y = -\xe e^{i k x}$. Note that with
an extra gauge transformation $\tilde \xi_x = i k y e^{i k x}$, we can get to
the metric in the normal coordinates used in \cite{Rosenhaus:2014woa} (which has $G_{xy}=0,\ G_{yy} \propto y K^y_{xx})$. This vector $\tilde \xi$ vanishes on the entangling surface ($y = 0$) and therefore the diffeomorphism it generates does not change the entanglement entropy. Substituting in \er{inducedmetric} and \er{metricperturb}, we get the leading order correction to entanglement entropy,
\ba
\Delta S_{EE} && = \int d^3x' \br  T_{xy}{}^{\CO(\xe)}(x'_{\mu}) \ke \delta G^*_{xy} (x'_{\mu}) \nonumber \\
&&=  \xe\tilde \xe \int d^3x' \int   d x dz \lsb 6ik \hat x z\left(-\frac{2}{\pi^2 f^5}2 (k z+1)\left(t'^2+\hat x^2-7 y'^2+z^2\right) \rc\rc \nn
& & \lc \lc-\frac{1}{\pi^2 f^4} k^2   \left(t'^2+\hat x^2-y'^2+z^2\right) \right)+\frac{4 k^{2} z (k z+1)}{\pi ^2 f^3}\rsb   e^{ik \hat x-kz} +\CO(\xe^3)\,.\qquad\quad
\ea
Note that $y(x,z)$ in $h_{zi}$ only gives contribution of higher orders in $\xe$ and therefore can be set to be zero. On the other hand, unlike $h_{zi}$, $h_{zz}$ is not multiplied by $\xe$ and therefore we need to expand $y\sim \CO(\xe)$ in $f$ and $C_{xy}$. The integration of $y'$ and $t'$ can be performed straightforwardly and the result is
\be
\Delta S_{EE} =\xe\tilde \xe \int dx' \int  dx dz\, k^2 z \lb
\frac{ -2ik \hat x }{\pi  \left(\hat x^2+z^2\right)^2}+\frac{ 2(k z+1) }{\pi  \left( \hat x^2+z^2\right)^2}\rb   e^{ik \hat x-kz}  \,.
\ee
We notice that $h_{zz}$ actually has no contribution to $\xd S_{\zt{EE}}$. The integration of $x$ is essentially the Fourier transform and we get
\ba
\Delta S_{EE} & = &   \xe \tilde \xe k^2 \int_\mu^\infty  dz  e^{-2 k z}\lb k^2+ \frac{ (k z+1)^2 }{ z^2}\rb \int dx'
\nn
& = &\xe \tilde \xe k^2 {(1+k\mu) e^{-2k\mu} \over \mu} \int d x'
= 2\xe \tilde \xe\lb \frac{k^2}{2 \mu }-\frac{k^3}{2}+\CO(\mu)\rb \int d x' \,,
\ea
where $\mu$ is the IR cutoff in the bulk. The result exactly reproduces \er{perturbedEEholo}.
To be more precise, we perform the replacement $\xe\to \xe + \tilde \xe$ in \er{perturbedEEholo} and find the terms of the order of $\CO(\xe \tilde \xe)$ agree.

\section{Conclusion}\label{conclude}
 
In dimensions higher than two, a co-dimension two twist operator is a surface operator and the precise formulation of its OPE with some local conserved charge density such as stress tensor have not been fully explored. As a starter, in this paper we study the OPE of stress tensor with a twist line operator in 3 dimensions.  We try an ansatz for such OPE coefficients constructed out of local geometric data encoding the shape of the line operator.  It turns out that the resulting stress tensor in the presence of the twist operator can not be conserved for general (wavy) shapes of line operator except for highly symmetric one such as planar or spherical  line operators. This is in contrast to the conserved holographic stress tensor, which we have constructed in this paper in the presence of the twist line operator based on RT formulation. Furthermore, we demonstrated that the stress tensor obtained holographically is not expressible as local geometric expansion, implying that the OPE of stress tensor with line twist operator cannot be expressed in terms of local geometric data only. It is possible that the Ward identities could be anomalous in the presence of surface operators, however, the holographic results suggest otherwise.

The perturbative computation we perform in sec~\ref{holo} is done holographically. It would be interesting to see whether the same stress tensor and entanglement entropy can be obtained from the integrated correlation functions of the CFT side using the second order perturbation (two insertions of stress tensors, see \eg \cite{Rosenhaus:2014woa}). However, it should be noted that there might be some subtleties in pushing the perturbative computation to second order. For example, it remains unclear \cite{Rosenhaus:2014zza} how to reproduce the shape dependence of the entanglement entropy (in four dimensions) using this approach.

 We should emphasize the failure of OPE in terms of local geometric data is peculiar in odd dimensions. In contrast, the universal part of the entanglement entropy in four dimensions is known to have local expressions \cite{Solodukhin:2008dh,Schwimmer:2008yh} in terms of (Graham-Witten) conformal anomalies encoding shape information of the twist operator. Thus, on hindsight, our results are not unexpected as there is no conformal anomaly in odd dimensions so that the universal part of the entanglement entropy has no even dimensional analogue. Thus, it deserves further study of the issues raised in the current paper to get a clearer understanding of the results in both even and odd dimensions.

\acknowledgments
LYH is grateful to inspiring discussions with R. Myers and J. Gomis and particularly S. Matsuura who collaborated at the early state of the project. FLL is supported by MoST grant:103-2112-M-003 -001 -MY3  and 103-2811-M-003 -024,  he is also supported by NCTS north and thanks for discussions with Juinn-Wei Chen. The work of XH is supported by MoST grant: 103-2811-M-003-024.

\appendix
\section{Useful Identities regarding the local geometrical quantities}\label{appendixA}
To obtain further expansions, we would need a few more building blocks.

To begin with, we need a complete basis of vectors in the vicinity of $\hat{x}$, let us also define the vector 
\be
k_\mu = \epsilon_{\mu\nu\rho} \hat{n}_\nu t_{\rho}.
\ee
One can readily check that $k_\mu k^\mu=1$ and by construction it is normal to both $\hat{n}_\mu$ and $t_\mu$.

We can evaluate the extrinsic curvatures, defined as
\be
K^a = - t^\mu t^\nu \nabla_{\mu} n^a_\nu,
\ee
where $n^a, a=\{1,2\}$ denotes the two possible unit normals $n^1= \hat{n}$ and $n^2= k$ respectively.

One very useful identity is
\be
\frac{\partial \sigma}{\partial x^\mu} = \frac{t_\mu}{1-\hat{n}^\alpha \partial_\sigma t_\alpha},
\ee
where $\sigma$ is a convenient parametrization of the entangling curve, such that it's trajectory in spacetime can be denoted by $\{y^\mu(\sigma)\}$, and that
\be
t^\mu = \mathcal{N}\frac{\partial y^\mu(\sigma)}{\partial \sigma},
\ee
where the normalization factor $\mathcal{N}=1$ if $\sigma$ is chosen to be some affine parameter.

The extrinsic curvature is therefore given by
\be
K^a = n^a_\nu \partial_\sigma t^{\nu},
\ee
for affine parameter $\sigma$. 

This identity is derived by requiring that as the insertion point $x^\mu$ is varied, $\hat{x}(\sigma_0)$ should vary accordingly so that 
\be
\hat{n}_\mu t^\mu=0
\ee
remains true.

There are a few useful identities:
\bea
\label{dn1}\partial_\mu n^1_\nu && = \frac{1}{|\epsilon|} \left( \delta_{\mu\nu} - t_\mu s_\nu - n^1_\mu n^1_\nu\right), \qquad s_\mu = \frac{\partial\sigma}{\partial x^\mu} = \frac{t_\mu}{1-|\epsilon|K^1}. \\
\label{dn2} \partial_\mu n^2_\nu && = \frac{1}{|\epsilon|}\xe_\nu{}^{\xa\xb} \left( \delta_{\mu\xa} t_\xb - t_\mu s_\xa t_\xb- n^1_\mu n^1_\xa t_\xb\right)+\xe_\nu{}^{\xa\xb} n^a_\xa K^a s_\mu n^1_\xb,\\
\partial_\mu |\epsilon| &&= n^1_\mu, \qquad \partial_\mu t^\mu = \frac{\partial^2 y^\mu}{\partial \sigma^2} s_\mu =0 \\
\partial_\mu K^a &&= \partial_\mu( n^a_\alpha \partial_\sigma t^\alpha)  = s_\mu K^b n^{a\,\,\beta} \partial_\sigma n^a_\beta + s_\mu \partial_\sigma K^a + K^b n^{b\,\,\beta} \partial_\mu n^a_\alpha \\
\label{dt1}\partial_{\mu}t^\nu &&= \frac{\partial^2 y^\nu}{\partial \sigma^2} s_\mu = n^a_\nu K^a s_\mu , \\
\frac{\partial t_\mu}{\partial \xs} &&= t^\alpha \partial_\alpha t_\mu = n^a_\nu K^a,\qquad \frac{\partial^2 t_\mu}{\partial \sigma^2} = t^\alpha \partial_\alpha t_\mu = \frac{n^a_\nu K^a}{1-|\epsilon| K^1}.\\
\partial_\xs n^a_\nu && =-K^a t_\nu,\qquad \p_\xs K^a =n^a_\alpha \partial_\sigma^2 t^\alpha 
\eea

\section{Holographic stress tensor in the presence of twist operator}\label{appendixB}
  Follow the same procedure in deriving $\br T_{xy} {}^{\CO(\xe)} \ke$, one can derive the other components of the stress tensor in the similar way. The final results are listed below by $k$-expansion and integrating out the worldvolume $\int dx \cdots$,
\ba
\br T_{xy} {}^{\CO(\xe)} \ke & = &\xe e^{i k x}\lsb \frac{3 i t^2 k}{8 \pi  \left(t^2+y^2\right)^{5/2}}-\frac{i \left(3 t^2+2 y^2\right)k^3}{16 \pi  \left(t^2+y^2\right)^{3/2}}-\frac{i
k^4}{8 \pi }+\CO(k^5)\rsb\,,\\
\br T_{ty} {}^{\CO(\xe)} \ke &= &\xe e^{i k x}\lsb -\frac{3 t \left(t^2-4 y^2\right)}{8 \pi  \left(t^2+y^2\right)^{7/2}} -\frac{3 t \left(t^2+2 y^2\right)k^2}{16 \pi  \left(t^2+y^2\right)^{5/2}} + \frac{t k^3}{4\pi (t^2+ y^2)} \rc \nn
 & & \lc -\frac{t \left(9 t^2+8 y^2\right)k^4}{64 \pi  \left(t^2+y^2\right)^{3/2}}+\CO(k^5)\rsb\,,\\
\br T_{yy} {}^{\CO(\xe)}\ke & = & \xe e^{i k x}\lsb \frac{3 y \left(y^2-4 t^2\right)}{8 \pi  \left(t^2+y^2\right)^{7/2}}-\frac{3 y^3 k^2}{16 \pi  \left(t^2+y^2\right)^{5/2}}+\frac{y
k^3}{4\pi (t^2+ y^2)}\rc \nn \label{Tyy0}
& & \lc -\frac{y \left(12 t^2+11 y^2\right)k^4}{64 \pi  \left(t^2+y^2\right)^{3/2}}+\CO(k^5)\rsb\,,
\ea
and
\ba
\br T_{xx} {}^{\CO(\xe)} \ke & = &\xe e^{i k x}\lsb \frac{3 y}{8 \pi  \left(t^2+y^2\right)^{5/2}}-\frac{3 y k^2}{16 \pi  \left(t^2+y^2\right)^{3/2}}+\frac{5 y k^4}{64 \pi  \sqrt{t^2+y^2}}+\CO(k^5)\rsb\,,\\
\br T_{tt} {}^{\CO(\xe)} \ke &= &\xe e^{i k x}\lsb \frac{3 \left(3 t^2 y-2 y^3\right)}{8 \pi  \left(t^2+y^2\right)^{7/2}}+\frac{3 y \left(t^2+2 y^2\right)k^2}{16 \pi  \left(t^2+y^2\right)^{5/2}}-\frac{y k^3}{4 \pi  \left(t^2+y^2\right)} \rc \nn
 & & \lc +\frac{y \left(7 t^2+6 y^2\right)k^4}{64 \pi  \left(t^2+y^2\right)^{3/2}}+\CO(k^5)\rsb\,,\\
\br T_{tx} {}^{\CO(\xe)}\ke & = & \xe e^{i k x}\lsb -\frac{3 i t y k }{8 \pi  \left(t^2+y^2\right)^{5/2}} + \frac{i t y k^3}{16 \pi  \left(t^2+y^2\right)^{3/2}}+\CO(k^5)\rsb\,,
\ea
where we have dropped the primes of $t'$ and $y'$ for simplicity.

\end{document}